\documentclass[aps,prd,reprint,superscriptaddress,nofootinbib,floatfix]{revtex4-1}

    \RequirePackage{amsmath}
    \RequirePackage{amssymb}
    \RequirePackage{graphicx}
    \RequirePackage[caption=false]{subfig}
    \RequirePackage{enumitem}
    \RequirePackage[usenames,dvipsnames]{color}
    \RequirePackage{mathtools}
    \RequirePackage[percent]{overpic}
    \RequirePackage{bm}
    \RequirePackage{rotating}
    \RequirePackage{cancel}
    \RequirePackage{hyperref}
    \RequirePackage{wrapfig}
    \RequirePackage[colorinlistoftodos]{todonotes}
    \RequirePackage{array}
    \RequirePackage[normalem]{ulem}
    \RequirePackage{ifthen}
    \RequirePackage{ifxetex}
    \RequirePackage{upgreek}

    \newcommand{\nocontentsline}[3]{}
    \newcommand{\tocless}[2]{\bgroup\let\addcontentsline=\nocontentsline#1{#2}\egroup}
    \ifthenelse{\isundefined{\showoldabstracts}}{
      \newcommand{\oldabstracts}[1]{}
    }{
      \newcommand{\oldabstracts}[1]{#1}
    }

    \def\shownal{1} 
    \newcommand{\nnaye}[1]{\ifthenelse{\shownal=1}{\textcolor{blue}{[[Naye: #1]]}}{}}
    \newcommand{\nemma}[1]{\ifthenelse{\shownal=1}{\textcolor{purple}{[[Emma: #1]]}}{}}
    \newcommand{\nedu}[1]{\ifthenelse{\shownal=1}{\textcolor{red}{[[Eduardo: #1]]}}{}}

    \newcommand{\ket}[1]{{\left| {#1} \right>}}
    \newcommand{\bra}[1]{{\left< {#1} \right|}}

    \newcommand{\ii}{\mathrm{i}}
    \newcommand{\ee}{\mathrm{e}}
    \newcommand{\tr}{\text{Tr}}

    \newcommand{\set}[1]{\ensuremath{\left\{\,#1\,\right\}}}
    
    \newcommand{\Ucorr}{\hat U_\text{corr}}
    \newcommand{\rhoA}{\hat\rho_\textsc{A}}
    \newcommand{\rhoB}{\hat\rho_\textsc{B}}

    \DeclareMathOperator{\sech}{sech}
    \DeclareMathOperator{\csch}{csch}
    \DeclareMathOperator{\setspan}{span}
    \DeclareMathOperator{\diag}{diag}

    \newcommand{\tensorupgreek}{
        \let\alpha\upalpha
        \let\beta\upbeta
        \let\chi\upchi
        \let\epsilon\upepsilon
        \let\varepsilon\upvarepsilon
        \let\eta\upeta
        \let\gamma\upgamma
        \let\omega\upomega
        \let\phi\upphi
        \let\psi\uppsi
        \let\sigma\upsigma
        \let\xi\upxi
        \let\zeta\upzeta
        \let\Lambda\Uplambda
        \let\Upsilon\Upupsilon 
        \let\Xi\Upxi
    }

\begin{document}

    \title{Fluctuations of work cost in optimal generation of correlations}

    \author{Emma McKay}
    \affiliation{Institute for Quantum Computing, University of Waterloo, Waterloo, Ontario, N2L 3G1, Canada}
    \affiliation{Department of Applied Mathematics, University of Waterloo, Waterloo, Ontario, N2L 3G1, Canada}

    \author{Nayeli A. Rodr\'{i}guez-Briones}
    \affiliation{Institute for Quantum Computing, University of Waterloo, Waterloo, Ontario, N2L 3G1, Canada}
    \affiliation{Department of Physics \& Astronomy, University of Waterloo, Waterloo, Ontario, N2L 3G1, Canada}
    \affiliation{Perimeter Institute for Theoretical Physics, 31 Caroline St. N., Waterloo, Ontario, N2L 2Y5, Canada}

    \author{Eduardo Mart\'{i}n-Mart\'{i}nez}
    \affiliation{Institute for Quantum Computing, University of Waterloo, Waterloo, Ontario, N2L 3G1, Canada}
    \affiliation{Department of Applied Mathematics, University of Waterloo, Waterloo, Ontario, N2L 3G1, Canada}
    \affiliation{Perimeter Institute for Theoretical Physics, 31 Caroline St. N., Waterloo, Ontario, N2L 2Y5, Canada}


    \begin{abstract}
    We study the impact of work cost fluctuations on optimal protocols for the creation of correlations in quantum systems. We analyze several notions of work fluctuations to show that even in the simplest case of two free qubits, protocols that are optimal in their work cost (such as the one developed by Huber et al. [NJP 17, 065008 (2015)]) suffer work cost fluctuations that can be much larger than the work cost. We discuss the implications of this fact in the application of such protocols and suggest that, depending on the implementation, protocols that are sub-optimal in their work cost could beat optimal protocols in some scenarios. This highlights the importance of assessing the dynamics of work fluctuations in quantum thermodynamic protocols.

    \end{abstract}

    \maketitle

    \section{Introduction}

    In the past decade, the study of the extension of thermodynamics to the quantum regime has seen significant advancements. From a quantum Landauer principle \cite{Funo2013,Goold2015,Sagawa2014} to resource theories and studies of allowable operations \cite{Alhambra2016a,Jevtic2012,Horodecki2013,Brandao2015}, the expanding field of quantum thermodynamics holds a great deal of insights. For detailed reviews, see e.g. \cite{Goold2016,Campisi2011}.

    In this context, the thermodynamics of quantum and classical correlations is a particularly interesting topic, especially within the framework of quantum information. Besides questions of how much entanglement can be generated in certain systems, what the fundamental limits are to entanglement generation, or what it can be used for, quantum thermodynamics has been addressing the energetic cost of generating correlations in quantum systems  \cite{Funo2013,Perarnau-Llobet2015,Jevtic2015,Huber2015,Bruschi2015,Friis2016}. For example, it has been shown that the average work cost of generating quantum correlations is larger than that of generating classical correlations \cite{Funo2013,Perarnau-Llobet2015}. Furthermore the generation (and destruction) of correlations under unitary action in particular was investigated in \cite{Jevtic2012,Jevtic2012a}, results that are especially relevant to this work.

    Within the topic of the energetics of correlations we find  the specific study of the work cost of creating bipartite correlations. Reminiscent of Landauer's principle, a lower bound on the average work cost $W$ of creating an amount of mutual information $I$ in a bipartite non-interacting system, initially in a thermal state of inverse temperature $\beta_\text{in}$, was derived in \cite{Bruschi2015}:
    \begin{equation} \label{eq:bruschi-ineq}
        \beta_\text{in} W \geq I.
    \end{equation}
    This bound cannot be saturated with unitary protocols; however, a protocol which does saturate this bound, developed in \cite{Bruschi2015}, makes crucial use of a correlating unitary developed in \cite{Huber2015} (and expanded on in \cite{Vitagliano2018}). This correlating unitary, in fact, creates maximal bipartite correlations at an optimal average energy cost.

    However, many studies have shown that the notion of \textit{average work} does not tell the whole story when it comes to the energetics of a protocol. The field of stochastic thermodynamics and the development of fluctuation relations discuss in detail the nature of work in a quantum setting \cite{Campisi2011,Seifert2012,Halpern2015,Salek2017,Dahlsten2017,Barra2017}. Not only are the average and the second moment (i.e. the fluctuations) important, but often even higher moments of work are nontrivial \cite{Dorner2013,Mazzola2013,Talkner2007}. These considerations include the study of the relationship of dissipated work (indicating irreversibility) to the presence of correlations, indicating that, at least in some cases, correlations make considering the distribution of work all the more important \cite{Carlisle2014,Galve2009}. The importance of considering fluctuations is made very clear by, for instance, the result that even in a thermodynamic limit, finite size systems retain a non-zero probability of transitions forbidden by the available average work \cite{Alhambra2016}.

    With this in mind, we explore in this paper the impact of quantum work fluctuations on the cost of generating correlations. In particular, we will analyze the unitary protocol creating maximal correlations with minimum work cost developed by Huber \textit{et al.} in \cite{Huber2015} for the case of two qubits.

    Remarkably, even though the importance of distributions and moments of work have been analyzed in the literature, the exact definition of the fluctuations of quantum work is not free of contention \cite{Talkner2016,Gallego2016,Perarnau-Llobet2017}. It is not trivial to decide which random variable describes work, nor how it emerges from quantum theory and thermodynamics. A number of options have been explored and justified for various purposes, including the widely used two-time energy measurement framework \cite{Talkner2007,Kurchan2000} and the ``work operator'' \cite{Yukawa2000}. 
    In the case we study below, the measures of fluctuations corresponding to these two conceptualizations of work are identical. We consider both this measure and a measure of the spreading of the energy distribution caused by the correlation-creating protocol, and discuss physical interpretations of these measures in some detail.

    We will thus show that the unitary protocol in \cite{Huber2015} suffers from large work cost fluctuations regardless of what metric of fluctuations we choose to evaluate them. This has implications in the adequacy of the protocol in scenarios where, for instance, one has a strict energy budget per run of the protocol, or where fewer than some number of runs of the protocol are possible. The work cost fluctuations are already of the order of the average work cost for low initial temperatures, and increase as the initial temperature increases. This poses significant issues for the reliability of the optimal nature of the protocol.

    In this aim, we first in section \ref{sec:protocol} detail the protocol as in \cite{Huber2015} and write explicitly the correlating unitary used. In section \ref{sec:noninteracting}, we give a thorough analysis of the two-qubit system and the effect of enacting protocol on it, including the average work cost. We then give some background on choices of random variables representing work in section \ref{sec:fluctuations}, and calculate two measures of uncertainty in the work cost. We show in the same section that the ratio of fluctuations to average work cost for optimal generation of mutual information can be of order 1, when beginning in a low-temperature (i.e. low-correlations) state, and that this ratio increases with the temperature of the initial state. We finish with conclusions and discussion in section \ref{sec:conclusion}.

    \section{Correlation-generating protocol and unitary of Huber \textit{et al.}} 
        \label{sec:protocol}

    We are interested in the single-step optimal correlation-generating protocol of \cite{Huber2015}. The protocol is optimal in the sense that it maximizes the amount of mutual information generated at the lowest possible average energy cost for a unitary protocol. It is not optimal in the sense of saturating the bound \eqref{eq:bruschi-ineq}---this is only achieved by using non-unitary cooling steps in addition to this unitary, as in \cite{Bruschi2015}. This protocol achieves optimal correlation generation by taking global thermal states to local Gibbs states with higher local temperatures \cite{Jevtic2012,Jevtic2012a}. For a non-interacting bipartite system, a global thermal state of inverse temperature $\beta_\text{in}$ is the Gibbs state 
        \begin{align} \label{eq:localthermal}
            \hat\tau(\beta_\text{in})=\frac{\ee^{- \hat H \beta_\text{in}}}{\tr [\ee^{- \hat H \beta_\text{in}}]}
        \end{align}
    This is a product state (i.e., completely uncorrelated) since the Hamiltonian is just the sum of two local terms. The single step of the protocol consists of the application of a bipartite correlating unitary $\Ucorr$ (on which we will elaborate below). After this step, the system is no longer in a global thermal state, acquiring correlations at the minimum work cost. In this section we will perform a relatively in-depth review of this protocol. 

    In \cite{Huber2015}, it is argued that for unitary protocols which generate correlations in non-interacting systems, target final states which have local Gibbs states (i.e., the subsystems are Gibbs states with respect to their local Hamiltonians) have the maximum possible mutual information $I$. Note the distinction between a thermal and Gibbs state; the Gibbs distribution, as in equation \eqref{eq:localthermal}, is that which maximizes entropy for a given inverse temperature $\beta_\text{in}$ and Hamiltonian $\hat H$. Thermal states are Gibbs distributions reached via a thermalization process. The application of the unitary here maximizes the local entropy on each subsystem, but does not do so through a thermalization process. For a bipartite quantum system $\hat \rho$ so that the partial subsystems' density matrices are denoted as $\rhoA\coloneqq\text{Tr}_\textsc{B}(\hat\rho)$, $\rhoB\coloneqq\text{Tr}_\textsc{A}(\hat\rho)$ the mutual information is given by
    \begin{equation}
        I(\hat\rho) \coloneqq S[\rhoA] + S[\rhoB] - S[\hat\rho],
    \end{equation}
    where $S[\hat\rho]=-\tr [\hat\rho\log\hat\rho]$ is the von Neumann entropy. $S[\hat\rho]$ will not change under application of the unitary and the entropies $S[\rhoA]$ and $S[\rhoB]$ are maximal (for a fixed average energy) for Gibbs states, so we indeed have that the mutual information will be maximal at a given expected energy when the partial states have Gibbs distributions, as claimed above.

    Huber \textit{et al.} prove (in Appendix A1 of \cite{Huber2015}) the existence of a unitary which transforms a $d^2$-dimensional bipartite systems in a global thermal state to a local Gibbs state. We review their proof in detail in Appendix \ref{app:proof}. In the body of their paper, they investigate the particular unitary which produces infinite temperature local Gibbs states. Here, we will look broadly at the class of unitaries which perform this function. We will call the family of unitaries which are shown to implement this transformation $\Ucorr$. 

    To see what this family of unitaries is, we need to make use of generalized Bell states.  Using the `clock' operators $\hat X$ and $\hat Z$ 
    \begin{align}
        \hat X &= \sum_{n=0}^{d-1}\ket{(n+1)\bmod d}\bra{n}, \\
        \hat Z &= \sum_{n=0}^{d-1}\ee^{2n \pi\ii /d}\ket{n} \bra{n},
    \end{align}
    we can write the generalized Bell states as
    \begin{align} \label{eq:bell-states}
        \ket{\phi_{m,n}} &= \hat Z^m \otimes \hat X^n \ket{\phi},
    \end{align}
    where $\ket{\phi}=d^{-1/2}\sum_{n=0}^{d-1} \ket{nn}$ and $d$ is the dimension of each subsystem (so that the dimension of the total system is $d^2$). We can write these states alternatively as
    \begin{align}
        \ket{\phi_{m,n}}&= \sum_{k=0}^{d-1} \ee^{2km\pi \ii/d} \ket{k} \otimes \ket{k+n},
    \end{align}
    which can be seen to be Bell states by noting that the label $n$ gives the difference between the state of the first and second system and the label $m$ gives the spacing of the phases on the $\ket{k}\otimes\ket{k+n}$.

    Now, these unitaries which transform global thermal states to local Gibbs states are those which perform rotations within the subspaces
    \begin{equation} \label{eq:general-subspace}
        S_i = \setspan \set{\ket{\phi_{0,i}},\ket{\phi_{1,i}},\dots, \ket{\phi_{d-1,i}}}.
    \end{equation}
    A review of the proof of this claim from \cite{Huber2015} can be found in Appendix \ref{app:proof}. The bases of these subspaces are those Bell states for which the difference between the label of the energy eigenstates of the first and second qubit is constant and equal to $i$, e.g. for two qubits, we have the zero difference subspace $S_0$ and the unit difference subspace $S_1$
    \begin{align} \label{eq:subspaces}
        S_0 &= \setspan\set{\ket{00}+\ket{11},\ket{00}-\ket{11}},\\
     \label{eq:subspaces2}   S_1 &= \setspan\set{\ket{01}+\ket{10},\ket{01}-\ket{10}}.
    \end{align}
    For a system of two qubits, then, the family of correlating unitaries $\Ucorr$ can be represented as a tensor product of two rotations $\hat R_0$ and $\hat R_1$ in the bases \eqref{eq:subspaces} and \eqref{eq:subspaces2} respectively:
    \begin{align} \nonumber
        \Ucorr &= \hat R_0 \otimes \hat R_1 \\
        &=
        \begin{pmatrix}
        \cos \theta     &   -\ee^{-\ii\delta}\sin\theta \\
        \ee^{\ii\delta}\sin\theta  &   \cos\theta
        \end{pmatrix}
        \otimes 
        \begin{pmatrix}
        \cos \phi     &   -\ee^{-\ii\gamma}\sin\phi \\
        \ee^{\ii\gamma}\sin\phi  &   \cos\phi
        \end{pmatrix},
    \end{align}
    where $\theta$, $\delta$, $\gamma$, and $\phi$ are real parameters that will be investigated in detail below. In the global energy eigenbasis  $\set{\ket{00}, \ket{01}, \ket{10}, \ket{11}}$, $\Ucorr$ takes the form
    \begin{widetext}
    \begin{equation}
    \displaystyle
        \label{eq:unitary}
        \Ucorr =
        \begin{pmatrix}
        \cos\theta+\ii \sin\delta\sin\theta & 0 & 0 & \cos\delta\sin\theta\\
        0 & \cos\phi+\ii \sin\gamma\sin\phi & \cos\gamma\sin\phi & 0\\
        0 & -\cos\gamma\sin\phi & \cos\phi-\ii \sin\gamma\sin\phi & 0 \\
        -\cos\delta\sin\theta & 0 & 0 & \cos\theta-\ii \sin\delta\sin\theta
        \end{pmatrix}.
    \end{equation}
    \end{widetext}

    Given this explicit form of the correlating unitary, we can not only implement the correlation-generating protocol, but also compute the relative magnitude of the work fluctuations as compared to the average work cost. In \cite{Huber2015} the optimization of work cost of creating correlations focuses only on the expected value. We will analyze the question of whether work cost fluctuations are actually relevant to decide how to optimize the energy cost of creating correlations.

    \section{Correlating qubits} 
        \label{sec:noninteracting}

    In this section we analyze both the average work cost and the work cost fluctuations of the protocol for the case of two qubits. 

    Consider a system of two identical qubits whose free Hamiltonian is
    \begin{equation}\label{eq:nonint-ham}
        \hat H=\frac\omega2\left( \hat\sigma_z^{(\textsc{a})}+ \hat\sigma_z^{(\textsc{b})}\right),
    \end{equation}
    so that the energy gap of both qubits is $\omega$. Let the two qubits be initially in a thermal state with temperature $\beta_\text{in}$. Thus, the initial state of the system is given by 
    \begin{equation}
        \hat\rho =\frac{\ee^{-\beta_\text{in} H}}{\tr[\ee^{-\beta_\text{in} H}]} 
        =\hat\tau_{_\textsc{A}}(\beta_\text{in})\otimes\hat\tau_{_\textsc{B}}(\beta_\text{in}),
    \end{equation}
    where $\hat\tau_{j}(\beta_\text{in})$ is the local Gibbs state of subsystem $j$.

    \subsection{Analysis of energy cost and correlations created for the protocol} 
        \label{sec:implementingprotocol}
    
    The initial state in the computational basis is represented by the thermal density matrix
    \begin{equation}
        \label{eq:inistate}
        \hat\rho=\frac{1}{Z}
        \begin{pmatrix}
        e^{-\beta_\text{in}\omega} & 0 & 0 & 0\\
        0 & 1 & 0 & 0\\
        0 & 0 & 1 & 0 \\
        0 & 0 & 0 &  e^{\beta_\text{in}\omega}
        \end{pmatrix},
    \end{equation}
    where $Z=\left(e^{-\beta_\text{in}\omega/2}+e^{\beta_\text{in}\omega/2}\right)^2.$ We now apply  the unitary  \eqref{eq:unitary} to the initial state: \mbox{$\hat\rho'=\Ucorr\hat\rho \Ucorr^\dagger$}. This yields
    \begin{equation}
        \label{eq:finstate}
        \hat\rho'=\frac{1}{Z}
        \begin{pmatrix}
        A_{-} & 0 & 0 & B_{+}\\
        0 & 1 & 0 & 0\\
        0 & 0 & 1 & 0 \\
        B_{-} & 0 & 0 &  A_{+}
        \end{pmatrix},
    \end{equation}
    where 
    \begin{align}
    \label{eq:A's}
    A&_{\pm} =e^{\pm\beta_\text{in}\omega}\mp 2 \operatorname{sinh}[\beta_\text{in}\omega]\operatorname{cos}^2\delta\operatorname{sin}^2\theta, \quad \rm{and} \\
     \label{eq:B's}
    B&_\pm=2\operatorname{sinh}[\beta_\text{in}\omega]\operatorname{cos}\delta\operatorname{sin}\theta\left(\operatorname{cos}\theta\pm \ii \operatorname{sin}\delta\operatorname{sin}\theta\right).
    \end{align}
    The initial symmetry of the state over the computational states $\ket{01}$ and $\ket{10}$ means that the parameters $\phi$ and $\gamma$ of the unitary have no effect on the final state of the system. The rotation in the subspace $S_1$ is irrelevant.
    
    \begin{figure*}
        \centering
        \includegraphics[width=0.8\textwidth]{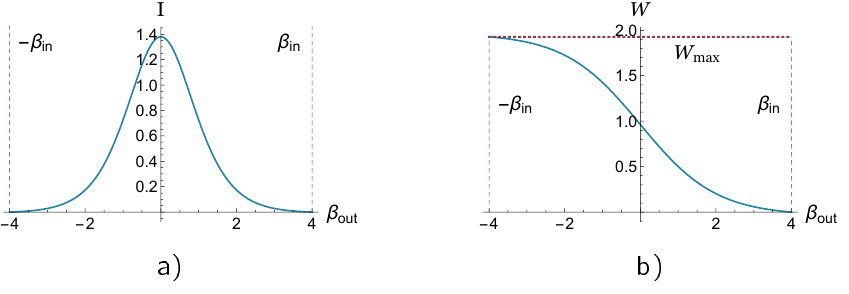}
        \caption{a) Mutual information $I$ created as a function of the final temperature $\beta_\text{out}$. The bounds on reachable $\beta_\text{out}$ are marked by dashed vertical lines at $-\beta_\text{in}$ and $\beta_\text{in}$. b) The work cost of the correlation-creating protocol as a function of the final temperature $\beta_\text{out}$. Bounds on the final temperature are marked by dashed vertical lines. The thick dotted horizontal line marks the upper bound on work cost, as in equation \eqref{eq:Workbound}.}
        \label{fig:work-beta-out}
    \end{figure*}

    The new local state of the qubits is obtained by partial tracing:
    \begin{align}
    \displaystyle
        \label{eq:finlostate}
        \hat\tau_{_\textsc{A}}(\beta_\text{out})&=
        \tr_{_\textsc{B}} [\hat\rho']=
        \frac{1}{Z}
        \begin{pmatrix}
        A_{-} +1 & 0 \\
        0 &  A_{+}+1
        \end{pmatrix},
    \end{align}
    and the state $ \hat\tau_{_\textsc{B}}$ has the same form in the local computational basis for the second qubit. $\hat\tau_{_\textsc{A}}(\beta_\text{out})$ and $\hat\tau_{_\textsc{B}}(\beta_\text{out})$  correspond to Gibbs states with respect to the local Hamiltonians $\hat H_{\textsc{a}/\textsc{b}}=\frac{1}{2}\omega\hat\sigma_z^{(\textsc{a})/(\textsc{b})}$ with inverse temperature
    \begin{equation}
        \beta_\text{out}=\frac{2}{\omega}\operatorname{arctanh}\left[\frac{A_+-A_-}{Z}\right].
    \end{equation}
    Substituting  eqs. \eqref{eq:A's} and \eqref{eq:B's} into this expression for $\beta_\text{out}$, we obtain that the initial and final inverse temperatures are related as follows
    \begin{equation} \label{eq:final_temp}
        \tanh{[\beta_\text{out}\omega/2]}=\tanh\left[\beta_\text{in}\omega/2\right]\left(1-2\cos^2{\delta}\sin^2{\theta}\right).
    \end{equation}
    Note that $\beta_\text{out}\in[-\beta_\text{in},\beta_\text{in}]$, with maximal creation of mutual information at $\beta_\text{out}=0$, as shown in Fig. \ref{fig:work-beta-out}.
    Even though $\Ucorr$ is a function of the parameters $\theta$, $\delta$, $\phi$, $\gamma$, the final temperature only depends on two of them, $\theta$ and $\delta$.
        
    The average internal energy  change after applying $\Ucorr$ to $\hat\rho$ will be given by the difference in energy expectation between the initial and final state:
    \begin{equation} \label{eq:Work}
        W=\tr[H(\Ucorr\hat\rho \Ucorr^{\dagger} -\hat\rho)].
    \end{equation}
    We denoted this with $W$ because, as the process is unitary, there is no average heat production, and the average energy change coincides with the average work cost.    
    
    
    We can find the explicit dependence of the expectation value of the work  cost (that we will shorten as \textit{average work cost}) on the parameters of the unitary and the initial temperature $\beta_\text{in}$ by combining \eqref{eq:unitary}, \eqref{eq:inistate}, and \eqref{eq:Work},
    \begin{equation} \label{eq:work-thetadelta}
        W(\beta_\text{in},\theta,\delta) = 2 \omega\tanh[\beta_\text{in}\omega/2] \cos^2 \delta \sin^2 \theta.
    \end{equation}
    The average work cost is upper-bounded by
    \begin{equation} \label{eq:Workbound}
        W\leq  2\omega\tanh[\beta_\text{in}\omega/2],
    \end{equation}
    corresponding to $\beta_\text{out}=-\beta_\text{in}$ and $I=0$.

    Note that the average work cost can also be easily written as a function of the initial temperature and final local temperature as follows:
    \begin{equation} \label{eq:Wtemps}
        W\left(\beta_\text{in}, \beta_\text{out},\omega \right)=\omega\left(\tanh[\beta_\text{in}\omega/2]-\tanh[\beta_\text{out}\omega/2]\right).
    \end{equation}

\begin{figure}[h]
    \centering
    \includegraphics[width=0.8\linewidth]{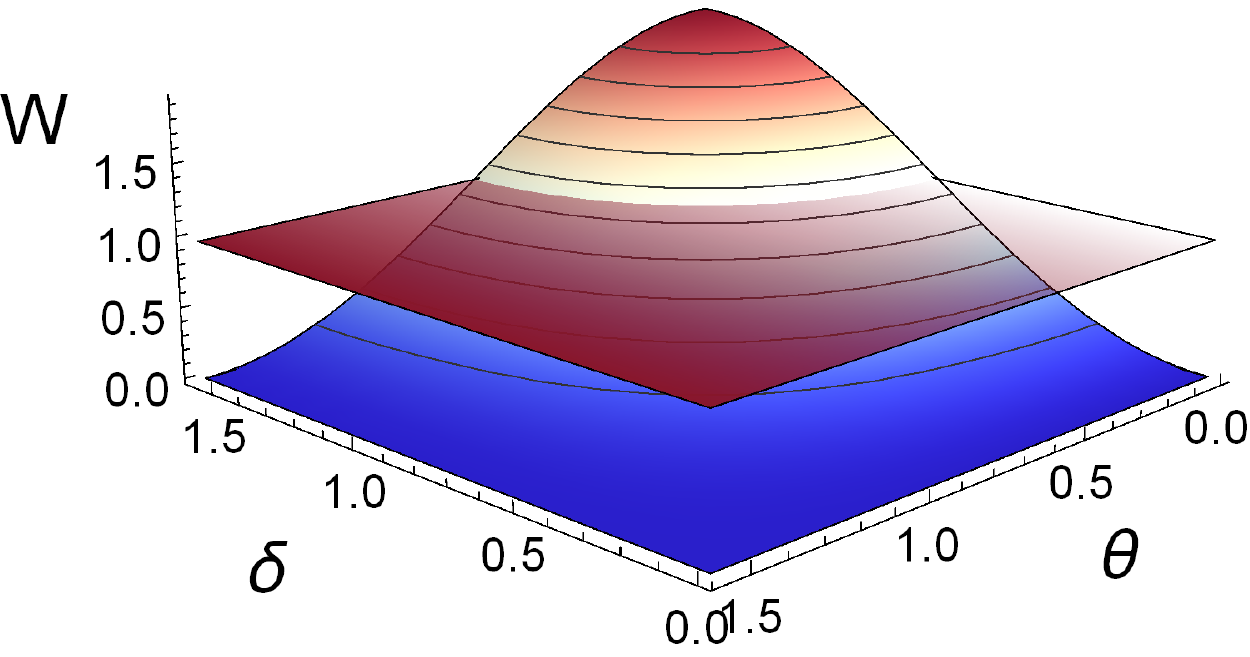}
    \caption{Average work cost as a function of the unitary parameters $\delta$ and $\theta$ for $\omega=1$, $\beta_\text{in}=4\omega^{-1}$. Lines of constant work as in equation \eqref{eq:thetadelta} are shown. The plane which cuts the plot identifies the work cost of maximal mutual information creation. Above this plane, the final local states have negative local temperature.}
    \label{fig:nonint-work-deltatheta}
    \end{figure}

    We see from \eqref{eq:work-thetadelta} that $W$ is independent of the parameters $\gamma$ and $\phi$ of the unitary operation. This implies that for a fixed value of $W$ we have an implicit relationship between $\theta$ and $\delta$. Hence, there are families of unitaries with the same average work cost that produce the same final local temperature (and the same amount of correlations), which are defined by the curves
    \begin{equation}\label{eq:thetadelta}
        \theta(\delta)= \arcsin \left[\sqrt{\frac{W \coth(\beta_\text{in} \omega/2)}{2\omega}} \sec \delta\right].
    \end{equation}
    These lines of constant average work cost can be seen in Fig.~\ref{fig:nonint-work-deltatheta}. 
    
    The amount of correlations that the application of the protocol creates between the qubits can be explicitly computed, and depends on the initial and final temperature as follows:
    \begin{equation} \label{eq:corr}
    \begin{split}
        I\left(\beta_\text{in}, \beta_\text{out},\omega\right)=\log&{\left[\frac{e^{\omega\beta_\text{in}\tanh{[\omega\beta_\text{in} /2]}}}{\cosh^2{[\omega\beta_\text{in}/2]}}\right]}\\
        &- \log{\left[\frac{e^{\omega\beta_\text{out}\tanh{[\omega\beta_\text{out} /2]}}}{\cosh^2{[\omega\beta_\text{out}/2]}}\right]}.
    \end{split}
    \end{equation}

    The unitary operation which gives the maximum amount of correlations between the two subsystems corresponds to the one which gives a final state with totally mixed local states. Indeed, the maximum amount of correlations possible is given by
    \begin{equation}
        I_{max}=\log{\left[\frac{e^{\beta_\text{in}\omega\tanh{[\beta_\text{in} \omega/2]}}}{\cosh^2{[\beta_\text{in}\omega/2]}}\right]}
    \end{equation}
    with an average work cost of 
    \begin{equation} \label{eq:maxcorr-work}
        W_{(I_{max})}=\omega \tanh{[\beta_\text{in}\omega/2]},
    \end{equation}
    corresponding to equation \eqref{eq:Wtemps} with $\beta_\text{out}=0$.

    \section{Quantifying work fluctuations} 
        \label{sec:fluctuations}

    \subsection{Work as a random variable}

    Classically, the notion of work is specifically the transfer of energy between a target system and an external system (often a type of battery, like a weight) produced through an active process. However, when the systems on which one performs work are quantum, the notion gets more challenging. To begin with, there is no clear observable (a self-adjoint operator) associated with work. This makes speaking of work fluctuations (and even the notion of work itself) tricky. 

    Here, we will provide background on two different random variables associated with work, and briefly recall that in the case we study, their associated measures of fluctuations are identical. We describe them both for completeness, and as a reminder of the current state of discussion around quantum work.

    The most popular definition of work as a random variable is currently the \textit{two-time energy measurement} framework. In this scheme, the work is the difference between final and initial energy eigenvalues found via projective measurements before and after application of the unitary \cite{Talkner2007,Kurchan2000}.

    This way of defining work as a random variable is inspired by classical stochastic thermodynamics, where there is a distribution of energy over a group of particles, but each particle nonetheless has a definite value of energy. At the beginning and end of a process, a particular particle may be chosen and measured; there is some definite and discrete energy difference in each particle. This is not the case for quantum systems, but the use of projective measurements emulates this notion, by ensuring the quantum system has a definite value of energy before and after the process. 

    Another notion of work as a random variable considers instead that there is inherent uncertainty in the energy of the system before and after the process. Indeed, we may not have access to projective measurements, they may be impossible to implement on our system, or we may not want to destroy the superposition in the energy eigenbasis. In this case, we consider the random variable representing work to be the difference of the random variables of the final and initial energies of the system. We can call this the \textit{non-projected work variable}. It has been discussed in the literature at several points, and is often called the work operator \cite{Allahverdyan2005,Talkner2007,Friis2018}.

    Conceptually, there are clearly big differences between these two random variables. It is even the case that the work associated with the two-time measurement scheme can be used to write a quantum Jarzynski equality, while the non-projected work cannot \cite{Talkner2007}. But there are a large class of cases when their first and second moments are actually the same---when $[\hat H,\hat\rho]=0$. Then, the average work cost of both schemes corresponds to equation \eqref{eq:Work}. If the Hamiltonian and the initial state do not commute, the average work of the two-time measurement scheme no longer corresponds to the differences in average energy of the final and initial states. Similarly, the fluctuations one measures under each scheme are identical when $[\hat H,\hat\rho]=0$. 


    \begin{figure*}
        \centering
        \includegraphics[width=0.9\textwidth]{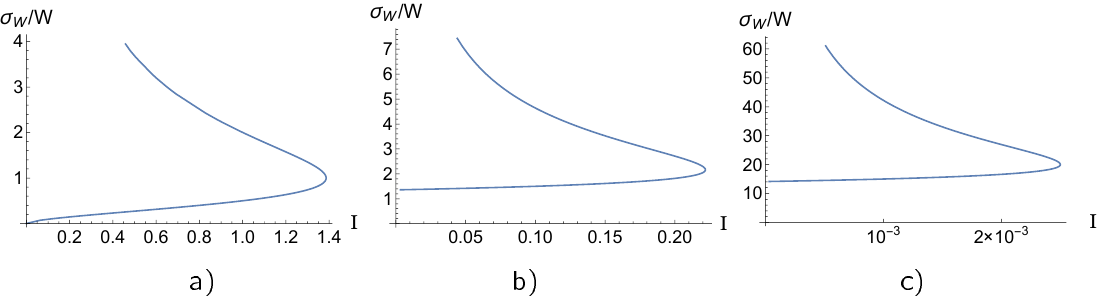}
        \caption{Ratio of work fluctuations, eq. \eqref{eq:literature-fluc}, to average work cost versus the change in mutual information with $\omega=1$, plotted parametrically as a function of increasing average work cost (from top to bottom of the curve) for three initial temperatures. From left to right, $\beta_\text{in} = 100\omega^{-1},\omega^{-1},0.1\omega^{-1}$. This ratio of fluctuations to average cost is of order unity for maximal creation of correlations for the lowest initial temperature, $\beta_\text{in}=100\omega^{-1}$, and increases with initial temperature. Note that the mutual information plotted here is that which is created in an application of the protocol without the two-time projective measurements. Work statistics of this protocol in a practical setting must be statistically generated over many implementations with projective measurements which destroy any correlations.}
        \label{fig:literature-fluc}
    \end{figure*}

    Let us write the details of these two conceptually different work schemes and their associated measure of fluctuations.
    In the two-time energy measurement framework, work is the difference between the final and initial energy found by projective measurement before and after the application of the unitary. The total probability distribution of this work $E_f-E_i$ is given by the probability of first measuring the system in an energy eigenstate $\ket{E_n}$ with eigenenergy $E_n$ and then transitioning to and measuring the system in eigenstate $\ket{E_m}$ with eigenenergy $E_m$, summed over all of these possibilities. For our unitary process $\Ucorr$, this probability distribution is
    \begin{equation}
        P(W) = \sum_{n,m} P(E_n) |\bra{E_m}\Ucorr\ket{E_n}|^2 \delta\big(W-(E_m - E_n)\big).
    \end{equation}
    The fluctuations $\sigma_{^W}$ in this probability distribution can be calculated as the square root of the variance, which for a probability distribution $P(W)$ is
    \begin{equation}
        \sigma_{^W}^2 = \int (W-W_\text{avg})^2 P(W) d W,
    \end{equation}
    or (see \cite{Allahverdyan2005}), since $[\hat\rho,\hat H]=0$, we can express it equivalently as the standard deviation in the operator \mbox{$\Delta \hat H = \Ucorr \hat H \Ucorr^\dagger - \hat H$},
    \begin{equation}
        \sigma_{^W} = \sqrt{ \tr(\Delta\hat H^2 \hat\rho) - \tr(\Delta \hat H \hat\rho)^2 }.
    \end{equation}
    Expanding the square of this quantity gives
    \begin{align}
        \sigma_{^W}^2 = \tr[&\hat H^2 \hat\rho] - \tr[\hat H \hat\rho]^2 + \tr[\hat H^2 \hat\rho'] - \tr[\hat H \hat\rho']^2 \\ \nonumber
        & - 2\big( \tr[\hat H \Ucorr^\dagger \hat H \Ucorr \hat\rho] - \tr[\hat H \hat\rho] \tr[\hat H \hat\rho'] \big),
    \end{align}
    which we may recognize as a combination of the variances in energy in the initial and final states and their covariance. We can write the work fluctuations as such,
    \begin{equation}
        \sigma_{^W}^2 = \sigma_{E_i}^2 + \sigma_{E_f}^2 - 2\sigma_{E_iE_f},
    \end{equation}
    where the energy variances are 
    \begin{align} \label{eq:energy-var}
        \sigma_{{E_i}}^2&=\tr[\hat H^2\hat\rho]-\tr[\hat H\hat\rho]^2,\\
        \sigma_{{E_f}}^2&=\tr[\hat H^2\hat\rho']-\tr[\hat H\hat\rho']^2
    \end{align}
    and the covariance is given by 
    \begin{equation}
        \begin{split}
        \sigma_{E_iE_f} &= \tr[\hat H \Ucorr^\dagger \hat H \Ucorr \hat\rho] - \tr[\hat H\hat\rho] \tr[\hat H \hat\rho']\\
        &= \langle \hat H (\Ucorr^\dagger \hat H \Ucorr) \rangle - \langle \hat H \rangle \langle \Ucorr^\dagger \hat H \Ucorr \rangle.
        \end{split}
    \end{equation}
    We have thus identified the fact that the work fluctuations are the same in all conceptual frameworks, dependent on $[\hat H,\hat\rho]=0$.
    
    In this protocol, the square of the fluctuations is
    \begin{align} \label{eq:literature-fluc}
        \sigma_{^W}^2 = & \frac{\omega^2 \cos^2\delta \sin^2\delta}{(1 + \ee^{\beta\omega})^2}
        \big(
        	3 + 2 \ee^{\beta\omega} + 3 \ee^{\beta\omega} \\ \nonumber
        	& \qquad\quad + (\ee^{\beta\omega} - 1)^2 (1 - 4 \cos^2\delta \sin^2\theta)
        \big).
    \end{align}
    In terms of the initial and final inverse temperatures, we can also write
    \begin{align}
        \sigma_{^W}^2 = \omega^2 
        \big( 
            \sech^2(\beta_\text{out}\omega/2) - 2\tanh(\beta_\text{out}\omega/2) \csch\beta_\text{in}\omega
        \big).
    \end{align}

    \begin{figure*}[ht]
    \centering
    \includegraphics[width=0.9\linewidth]{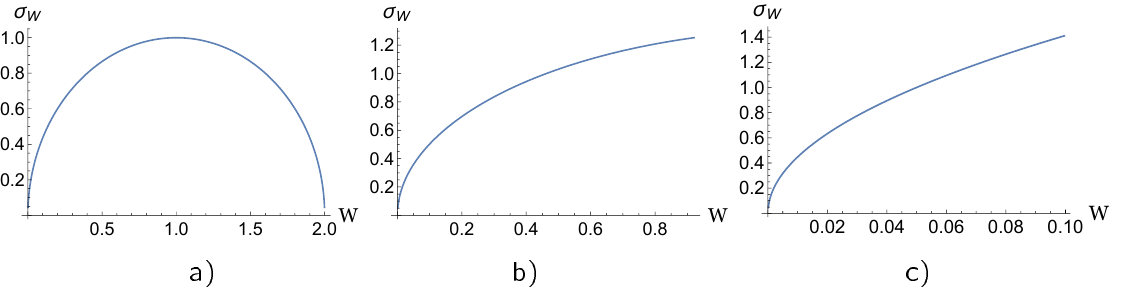}
    \caption{Work fluctuations $\sigma_W$ versus average work cost, for $\omega=1$ and initial temperatures, from left to right, of $\beta_\text{in}=100\omega^{-1}$, $\omega^{-1}$, and $0.1\omega^{-1}$.}
    \label{fig:flucs-v-work}
    \end{figure*}   

    The ratio of $\sigma_W$ to the average work cost is plotted in Fig.\ref{fig:literature-fluc} as a function of the change in mutual information for three different initial temperatures. We consider the production of mutual information and the generation of work statistics to be conducted in separate instances. We may implement the protocol to generate mutual information, but to collect work statistics in a practical setting, we must separately iterate the protocol many times, destroying any correlations created in the application of both projective measurements. We note that this ratio of fluctuations to average work cost is significant for the case of maximal creation of correlations---it is of order unity even for the lowest initial temperature investigated, and only increases with the initial temperature of the system. 

    The average work cost increases as we move along the curve from the top to the bottom, showing that the protocol, after reaching its peak mutual information generation, becomes increasingly less effective in creating correlations. Note, therefore, that the points of $I=0$ in the bottom of those plots do not correspond with doing nothing (identity process). Rather, they correspond to producing final local states with $\beta_\text{out}=-\beta_\text{in}$ i.e. those which saturate \eqref{eq:Workbound}. The protocol is not capable of generating local Gibbs states with temperature more negative than this.

    We plot the work fluctuations as a function of work in figure \ref{fig:flucs-v-work}. Note their increase in magnitude for higher initial temperatures, even as the maximum possible work cost decreases significantly.

  
    \subsection{Change in energy standard deviation}


    \begin{figure*}
        \centering
        \includegraphics[width=0.9\textwidth]{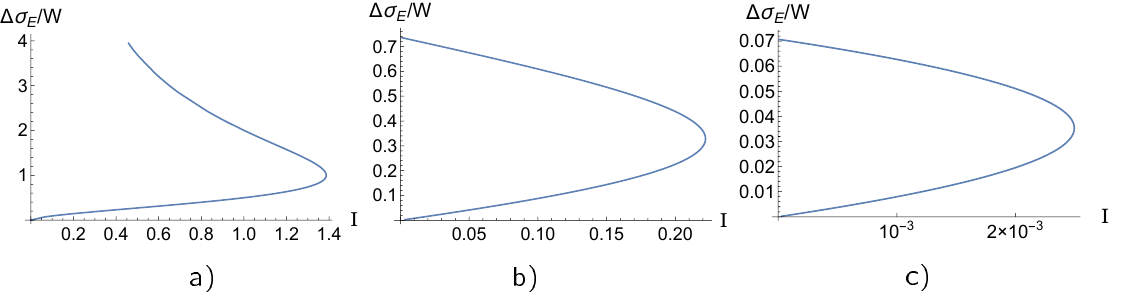}
        \caption{Ratio of change in energy standard deviation $\Delta\sigma_{_E}$ to work cost versus change in mutual information with $\omega=1$, plotted parametrically as a function of increasing average work cost (from top to bottom of the curve) for three initial temperatures. From left to right, $\beta_\text{in}=100\omega^{-1},\omega^{-1},0.1\omega^{-1}$. The magnitude of this ratio decreases significantly as the temperature of the initial state increases, suggesting that a large amount of the work fluctuations and work standard deviation is due to the standard deviation of the energy of the initial state, rather than due to the application of the unitary.}
        \label{fig:changevar}
    \end{figure*}

    Another question of interest about the protocol could be: by how much does the application of the protocol modify the intrinsic uncertainty in energy of the state? Such a quantity can provide an indication of the role that the protocol plays in the value of the work standard deviation (i.e. it would remove the contribution to the work standard deviation from the initial state) by looking at the change in standard deviation of the energy:
    \begin{align}
        \Delta \sigma_{_E} \coloneqq \sigma_{_{E_f}} - \sigma_{_{E_i}},
    \end{align}
    where the $\sigma_{_{E_i}}$ and $\sigma_{_{E_f}}$ are defined as in equation \eqref{eq:energy-var}. This change in standard deviation has substantially different behaviour than the work standard deviation and the work fluctuations, as shown in Fig.\ref{fig:changevar}. We see that the change in standard deviation $\Delta\sigma_{_E}$ is very similar to the total standard deviation $\sigma_{_W}$ when the initial temperature is low, as may be expected. The change in standard deviation decreases as the initial temperature increases, unlike the total standard deviation, indicating that a substantial amount of the uncertainty in the work cost of the unitary protocol is due more to the temperature of the state rather than to the application of the unitary itself. This does not mitigate the uncertainty in the actual work cost of applying the unitary. It does, however, caution us as to the existence of a protocol which will have lower uncertainty for higher temperature states.
 
    \section{Discussion and Conclusions} 
        \label{sec:conclusion}

    What is the impact of quantum fluctuations on the cost of generating correlations? To address this question, we have analyzed the protocol developed by Huber \textit{et al.} in \cite{Huber2015} for the case of two qubits.  This protocol was designed to generate maximal correlations at the minimal (average) work cost in unitary protocols on non-interacting quantum systems.

    Specifically, we have investigated the average work cost, the work cost fluctuations, and the mutual information generated by the protocol. 
    In keeping with the current discussion of how best to measure quantum work \cite{Baeumer2018}, we calculate the work cost fluctuations keeping the random variable we are taking to describe work in mind. In this case, the fluctuations corresponding to the two-time energy measurement and the work operator schemes are identical, and are used as our measure of fluctuations. 

    We have found that the optimal protocol developed in \cite{Huber2015} suffers work cost fluctuations that dwarf the average work cost of generating correlations even at low initial temperatures, and these fluctuations increase rapidly as the system's initial temperature increases. We find that the ratio of fluctuations in work cost to the average work cost is at least of order one for maximal correlation generation, even in low temperature initial states.

    We also investigated the change in standard deviation of the energy of the state as a consequence of applying the protocol. This measure provides an indication of the uncertainty in the work cost which is due only to the unitary and not to the uncertainty in the energy of the initial state. We find that this measure decreases significantly for initial states with high temperatures, indicating that a great deal of uncertainty may be at least heuristically due to the target states rather than the unitary protocol itself. Regardless of where they come from, work fluctuations are still present and significant regardless of the initial temperature of the state, and ought to be considered for practical implementation of such a unitary protocol.

    When fluctuations are much higher than expectation values, an analysis of only the average work cost may be insufficient in a number of scenarios. Fluctuations are relevant, for instance, if the energy budget per use of the protocol is limited. In those cases the actual single-shot work cost of the protocol can greatly exceed the budget. More importantly, this impacts the number of times that the protocol has to be applied for the statistical noise in the work cost to be irrelevant. In this case the work cost can quickly go orders of magnitude above the average energy cost of the protocol. Hence, in order for the expectation value of the work cost to give a good estimate of the total cost of creating correlations one needs to repeat the protocol a large (and rapidly increasing with temperature) number of times, or else the actual cost can be orders of magnitude off its expectation value.  

    If one cannot afford the luxury of repeating the protocol so many times, there may be other protocols that, while suboptimal in the sense of having higher average work cost than the unitary protocol given in \cite{Huber2015}, are nonetheless more reliable in terms of having smaller fluctuations. We wish to be cautious, however: the existence of such a protocol at high temperatures is potentially called into question by the fact that the change in energy standard deviation induced by application of the unitary is quite small at high temperatures. Regardless, even the simple case analyzed here already illustrates that to fully understand the work cost of a quantum protocol it is not enough to focus only on its expectation value.

    \acknowledgments

    The authors are very grateful to Marcus Huber, Karen Hovhannisyan, Marti Perarnau and Antonio Ac\'in for the truly illuminating discussions and feedback that shaped the final form of this article. E.M-M acknowledges support of the NSERC Discovery program and his Ontario Early Researcher Award. N.R-B acknowledges support of CONACYT.

    \appendix
    \section{Proof of the existence of the correlating unitary \texorpdfstring{$\Ucorr$}{U}}
        \label{app:proof}

    In this appendix, we provide a detailed restatement of the proof in Appendix A.1 of \cite{Huber2015}, which proves that the correlating unitary we use acts as we claim. Namely, we wish to show that, for a bipartite system of dimension $d^2$ composed of two subsystems A and B, of local dimension $d$, a unitary consisting of rotations in the subspaces spanned by the generalized Bell states \eqref{eq:bell-states},
    \begin{align}
        S_j = \operatorname{span} \set{\ket{\phi_{0,j}}, \ket{\phi_{1,j}}, \dots, \ket{\phi_{d,j}} },
    \end{align}
    will transform a global thermal state of inverse temperature $\beta_\text{in}$ to a local Gibbs state of some local inverse temperature $\beta_\text{out}$. Additionally, we will show that the unitary may be tailored to reach a specific final temperature.

    Under the assumption that there is no interaction Hamiltonian between the two parts of the system, we will work in the eigenbasis of the local Hamiltonians that we notate as $\{\ket{ij}\}$.

    Let us first summarize the steps of the proof. We will begin by showing that the action of rotating the initial global thermal state in the subspaces $S_j$ produces global coherences in the energy eigenbasis, but no local coherences. This proves that the action of the unitary takes global thermal states to locally diagonal states (a necessary condition for the local final states to be Gibbs). We will deduce, then, that the action of the global unitary on the local subsystems may be described by a particular type of transformation called circulant doubly stochastic transformation (CDST). We use this to rewrite the global unitary action on the subsystems in terms of a general CDST. Rewriting the general transformation in this way will allow for the calculation of the specific map needed to take a global thermal state of inverse temperature $\beta_\text{in}$ to a local Gibbs state of inverse temperature $\beta_\text{out}$.

    Let us briefly review notation. We call the initial global state $\hat\rho$. This is a global thermal state. The global state after the application of the unitary is $\hat\rho'$, generally not in a global Gibbs state. The local states of subsystem A and B are denoted $\rhoA$ and $\rhoB$, with primes similarly denoting the local states after application of the unitary. We notate matrix elements of $\hat \rho$ as follows
    \begin{align}\label{A3} 
        \hat\rho &= \sum_{i,j,k,l=0}^{d-1} \rho_{_{ij,kl}} \ket{ij}\bra{kl}.
    \end{align}
    Since there is some amount of index gymnastics in this appendix, when the value of an index gets too long we use a bracket notation like $\rho_{_{(d-1)(d-1),(d-1)(d-1)}}$ to separate the first and second indices and the third and fourth indices. Matrix elements of the state $\rhoA$ are denoted by $\alpha_{_{i,j}}$. They are given by partial tracing system B from the bipartite density matrix $\hat \rho$: 
    \begin{equation} \label{eq:partial-trace}
        \alpha_{_{i,j}} = \sum_{k=0}^{d-1} \rho_{_{ik,jk}}.
    \end{equation}

    Now, the action of the unitary on the global state is to mix elements within each subspace $S_j$. A particular rotation in a subspace $S_j$ will, in general, act on elements of the form
    \begin{equation} \label{eq:element-form}
        \rho_{_{n(n+j),m(m+j)}}.
    \end{equation}
    For example, consider rotations in the subspace $S_1$. This is the space spanned by the generalized Bell states
    \begin{align}
        \ket{\phi_{n,1}} = \sum_{k=0}^{d-1} \ee^{2k\pi\ii/d} \ket{k}\ket{k+1}.
    \end{align}
    The action of rotating in $S_1$  will only affect density matrix elements like $
        \rho_{_{01,01}},\,  \rho_{_{01,23}}, \, \rho_{_{23,12}},\,\dots,$
    or, in general, matrix elements of the form $
        \rho_{_{n(n+1),m(m+1)}}$.

    We wish to show that the off-diagonal elements of $\hat\rho'$ do not contribute to the partial trace, ensuring that the local states $\rhoA'$ and $\rhoB'$ maintain their diagonal nature, i.e., the unitary does not create coherences in the local states. To prove this we will use the observation made above. 

    Since the initial global state is thermal, it is diagonal in the Hamiltonian eigenbasis. That implies that matrix elements of the global state which are non-zero after application of the unitary are of the form $\rho'_{_{n(n+j),m(m+j)}}$, and matrix elements contributing to the local state $\rhoA'$ are of the form $\rho'_{_{ik,jk}}$. Thus, the matrix elements of the final global state which are both nonzero and contribute to the local state have \mbox{$n+j=m+j$}, i.e. $n=m$, that is, diagonal elements of $\hat\rho'$. Global diagonal matrix elements only contribute to local matrix elements which are also diagonal. Thus, the action of the unitary on a global thermal state does not create coherences in the local states. Hence, since the local states were initially diagonal (because the systems are non-interacting) then the local states remain diagonal after the application of the unitary.

    To continue the proof, let us introduce the notion of doubly-stochastic transformations. A matrix $T$ is doubly stochastic in a given basis if and only if its matrix elements satisfy
    \begin{align}
        T_{ij} &> 0\qquad \text{and}\\
        \sum_i T_{ij} &= \sum_{j} T_{ij} = 1.
    \end{align}
    A stochastic transformation of a vector preserves the sum of its components: considering $x=Ty$ for $x$ and $y\in\mathbb{R}^d$, we have that
    \begin{align}
        \sum_{i=0}^{d-1} x_i &= \sum_{i,j=0}^{d-1} T_{ij} y_j\\ \nonumber
        &= \sum_{j=0}^{d-1} y_j \left(\sum_{i=0}^{d-1} T_{ij} \right)\\ \nonumber
        &= \sum_{j=0}^{d-1} y_j.
    \end{align}
    In fact, all matrices which preserve the sum of vector elements satisfy $\sum_i T_{ij} = 1$ \cite{Ando1989}. The final property, $\sum_j T_{ij}=1$, is the defining characteristic of unital matrices, that is, matrices whose action preserves the `fully mixed' vector $[1,1,\dots,1]^\intercal$ \cite{Watrous2018}. 

    A global unitary transformation on $\hat \rho$ preserves its trace. All unitary transformations are also unital.
    This implies that the diagonals of the global state undergo a doubly stochastic transformation, that is,
    \begin{align}
        \diag (\hat\rho') = T \diag(\hat\rho),
    \end{align}
    where $\diag(\hat\rho) = (\rho_{_{00,00}},\rho_{_{01,01}}...)^\intercal$ and $T$ is a doubly stochastic matrix. This will be useful for the rest of the proof.

    Indeed, we can further visualize the action of the unitary through the fact that the rotations in each subspace $S_j$ only act on diagonal matrix elements $\rho_{_{n(n+j),n(n+j)}}$, $n=0,\dots,d-1$. We thus have that the unitary implements doubly stochastic transformations on each set of diagonal matrix elements belonging to different subspaces $S_j$:
    \begin{align}
        \begin{bmatrix}
        \rho'_{_{0j,0j}}\\
        \rho'_{_{1(1+j),1(1+j)}}\\
        \vdots\\
        \rho'_{_{(d-1)(d-1+j),(d-1)(d-1+j)}}
        \end{bmatrix}
        =
        T^{(j)}
        \begin{bmatrix}
        \rho_{_{0j,0j}}\\
        \rho_{_{1(1+j),1(1+j)}}\\
        \vdots\\
        \rho_{_{(d-1)(d-1+j),(d-1)(d-1+j)}}
        \end{bmatrix},
    \end{align}
    where $T^{(j)}$ is a doubly stochastic matrix corresponding to the elements belonging to the subspace $S_j$.

    For convenience, let us write a shorthand for the above vectors:
    \begin{align}
        \diag(\hat\rho)_{_{S_j}} = 
        \begin{bmatrix}
        \rho_{_{0j,0j}}\\
        \rho_{_{1(1+j),1(1+j)}}\\
        \vdots\\
        \rho_{_{(d-1)(d-1+j),(d-1)(d-1+j)}}
        \end{bmatrix}.
    \end{align}
    We can actually write the partial trace by summing over these vectors, as
    \begin{align}
        \diag(\rhoA) = \sum_{j=0}^{d-1} \diag(\hat\rho)_{_{S_j}}.
    \end{align}
    To see this, compare the elements of the vectors $\diag(\hat\rho)_{_{S_j}}$ with the partial trace as in \eqref{eq:partial-trace}.

    Then the action of the unitary on the local state can be finally written as 
    \begin{align}
        \diag(\rhoA') = \sum_{j=0}^{d-1} T^{(j)} \diag(\hat\rho)_{_{S_j}}.
    \end{align}

    The action on the second subsystem can be expressed in terms of the same doubly stochastic matrices $T^{(j)}$ by making use of the permutation operator $\Pi=\sum_k \ket{k}\ket{k}\bra{k+1}\bra{k+1}$, which shifts coefficients $\rho_{_{ij,kl}}$ to apply to different elements as: \mbox{$\Pi \rho_{_{ij,kl}}\ket{ij}\bra{kl} \Pi^\dagger = \rho_{_{ij,kl}}\ket{(i+i)(j+1)}\bra{(k+1)(l+1)}$}. For the rest of this paper, we will employ a shorthand notation to use with the vector $\diag(\hat\rho)_{_{S_j}}$ as,
    \begin{align}
        \Pi^j \diag(\hat\rho)_{_{S_j}} = 
        \begin{bmatrix}
        \rho_{_{(-j)0,(-j)0}}\\
        \rho_{_{(1-j)1,(1-j)1}}\\
        \vdots\\
        \rho_{_{(d-1-j)(d-1),(d-1-j)(d-1)}}
        \end{bmatrix},
    \end{align}
    where all indices are taken to be $\!\!\mod d$, so that, for instance $\rho_{_{(-1)0,(-1)0}}=\rho_{_{(d-1)0,(d-1)0}}$.
    Then the diagonals of subsystem B can be written as
    \begin{align}
        \diag(\rhoB) = \sum_{j=0}^{d-1} \Pi^j \diag(\hat\rho)_{_{S_j}}.
    \end{align}
    We thus write the action of the unitary on the second subsystem as
    \begin{align} \label{eq:diag-rho-b}
        \diag(\rhoB') = \sum_{j=0}^{d-1} \Pi^j (T^{(j)} \diag(\hat\rho)_{_{S_j}}).
    \end{align}

    If the action of the specified unitary is symmetric on both subsystems (guaranteed in our case because the local Hamiltonians of both qubits are the same \cite{Vitagliano2018}), this directly implies that the doubly stochastic matrices $T^{(j)}$ commute with the permutation operator $\Pi$, i.e., the $T^{(j)}$ are circulant. To find a particular transformation that allows us to set the temperature of the local states to any arbitrary $1/\beta_\text{out}$ it is enough to pick all the rotations on the different subspaces $S_j$ to have the `same angle', that is to take all the $T^{(j)}$ to have the same matrix elements to one another. We can then write
    \begin{align}
        \diag(\tr_\textsc{B}[\hat\rho']) &= T \sum_{j=0}^{d-1} \diag(\hat\rho)_{_{S_j}},\\
        \diag(\tr_\textsc{A}[\hat\rho']) &= T \sum_{j=0}^{d-1} \Pi^j \diag(\hat\rho)_{_{S_j}}.
    \end{align}
    We can write $T$ (or any doubly stochastic matrix) in terms of a convex combination of permutation matrices (Birkhoff-von~Neumann decomposition \cite{Dufosse2016}), the coefficients of which can be calculated from the initial and final temperatures of the system. Concretely,
    \begin{align} \label{eq:birkhoff-vonneumann}
        T = \sum_{i=0}^{d-1} \eta_i \Pi^i,
    \end{align}
    where the $\eta_i$ satisfy $\eta_i>0\,\forall\,i$ and $\sum_i \eta_i =1$.
    The diagonal elements of the local states after application of the unitary will thus be given by
    \begin{align} \label{eq:a20}
        \diag(\rhoA') &= \sum_{i=0}^{d-1} \eta_i \Pi^i \sum_{j=0}^{d-1} \diag(\hat\rho)_{_{S_j}},\\
        \diag(\rhoB') &= \sum_{i=0}^{d-1} \eta_i \Pi^i \sum_{j=0}^{d-1} \Pi^j \diag(\hat\rho)_{_{S_j}}.
    \end{align}
    Let us again focus on the state of subsystem A to determine the coefficients $\eta_i$ from a given initial and desired final temperature. We can rewrite the right hand side of \eqref{eq:a20} as
    \begin{widetext}
    \begin{align} \nonumber
        \sum_{i=0}^{d-1} \eta_i \Pi^i 
        \begin{bmatrix}
        \alpha_{00}\\
        \alpha_{11}\\
        \vdots\\
        \alpha_{(d-1)(d-1)}
        \end{bmatrix}
        &= 
        \eta_0 
        \begin{bmatrix}
        \alpha_{00}\\
        \alpha_{11}\\
        \vdots\\
        \alpha_{(d-1)(d-1)}
        \end{bmatrix}
        +
        \eta_1
        \begin{bmatrix}
        \alpha_{(d-1)(d-1)}\\
        \alpha_{00}\\
        \vdots\\
        \alpha_{(d-2)(d-2)}
        \end{bmatrix}
        +
        \eta_2
        \begin{bmatrix}
        \alpha_{(d-2)(d-2)}\\
        \alpha_{(d-1)(d-1)}\\
        \vdots\\
        \alpha_{(d-3)(d-3)}
        \end{bmatrix}
        + \cdots
        \\
        &=
        \begin{bmatrix}
        \eta_0 \alpha_{00} + \eta_1 \alpha_{11} + \cdots\\
        \eta_0 \alpha_{11} + \eta_1 \alpha_{22} + \cdots\\
        \eta_0 \alpha_{22} + \eta_1 \alpha_{33} + \cdots\\
        \vdots\\
        \eta_0 \alpha_{(d-1)(d-1)} + \eta_1 \alpha_{00} + \cdots
        \end{bmatrix},
    \end{align}
    \end{widetext}
    so that we can write each new diagonal element of the local state as
    \begin{align}
        \alpha'_{jj} = \sum_{i=0}^{d-1} \eta_i \alpha_{(i+j)(i+j)}.
    \end{align}
    Recalling, again, that both the final and initial local states are Gibbs allows us to rewrite initial and final diagonal matrix elements of the local states as
    \begin{align} \nonumber
        \frac{\ee^{-\beta_\text{out} E_j}}{Z'} &= \sum_{i=0}^{d-1} \eta_i \frac{\ee^{-\beta_\text{in} E_{i+j}}}{Z}\\
        \label{eq:eta}
        \Rightarrow
        \ee^{-\beta_\text{out} E_j} &= \frac{Z'}{Z} \sum_{i=0}^{d-1} \eta_i \ee^{-\beta_\text{in} E_{i+j}},
    \end{align}
    where we have defined the partition functions \mbox{$Z=\tr[\ee^{-\ii\beta_\text{in} \hat H}]$} and $Z'=\tr[\ee^{-\ii\beta_\text{out} \hat H}]$. This establishes a simple linear relationship between all the $\eta_i$ and the final and initial temperatures that can easily be inverted to find all possible transformations $\eta_i(\beta_\text{in},\beta_\text{out})$ (the solution is not unique). Not all collections of coefficients determined by pairs of temperatures $\beta_\text{in}$ and $\beta_\text{out}$ yield valid transformations (i.e., if the $\eta_i$ compatible with \eqref{eq:eta} were negative). This means that there may be some sets of initial and final temperatures for which the unitary transformation that we desire is not possible. For Hamiltonians with uniform energy spacings, the final temperature is limited to be in the range $[-\beta_\text{out},\beta_\text{out}]$. For Hamiltonians with nonuniform energy spacings, this range is further constricted. There is, then, a minimum local temperature that the application of this unitary can produce; beginning in a global thermal state of inverse temperature $\beta_\text{in}$, we must have $\beta_\text{out}<\tilde\beta$ for some $\tilde\beta<\beta_\text{in}$ \cite{Vitagliano2018}. This is not a significant roadblock since then this transformation is sufficient to reach any $\beta_\text{out}<\tilde\beta$.

    This completes the proof. In summary, we have presented here a pedagogical summary of the result in Appendix A.1 of \cite{Huber2015}. Repeating the steps in \cite{Huber2015}, we have shown that the unitary which performs rotations in the subspaces \eqref{eq:general-subspace} transforms global thermal states to states which have local Gibbs distributions. We have shown how the protocol in \cite{Huber2015} gives a method to calculate the transformation acting on the local state, assuming that the rotation in each subspace $S_j$ is the same. 

    Notice that there is a relevant difference between the method used in the body of the paper and the method from \cite{Huber2015} that we summarized in this appendix. The latter does not fully specify the global unitary, only the local transformations.  In the body of the paper, we use the general form for the global unitary that yields local transformations as the ones desired in this appendix. We additionally do not assume that the rotations in each subspace are the same, although we found that one of the rotations is indeed irrelevant. 

\bibliography{bibliography}
\end{document}